\documentclass[useAMS,usenatbib]{mn2e}
\usepackage{aasabbrev}
\usepackage{graphicx}
\usepackage{subfigure}
\usepackage{nicefrac}
\usepackage{color} 
\def\ltsim{\lower.5ex\hbox{$\; \buildrel < \over \sim \;$}}
\def\gtsim{\lower.5ex\hbox{$\; \buildrel > \over \sim \;$}}
\newcommand{\Msun}{{\rm M}_\odot}
\newcommand{\rs}{r_{\rm s}}
\title[A cosmological context for compact massive galaxies]{A cosmological context for compact massive galaxies}
\author[M. J. Stringer et al.]{\parbox{\textwidth}{Martin Stringer\thanks{mstringer@iac.es}, Ignacio Trujillo, Claudio Dalla Vecchia \\
and Inma Martinez-Valpuesta}\vspace{0.5cm}\\
\parbox\textwidth{Instituto de Astrof\'{i}sica de Canarias, E-38205 La Laguna, Tenerife, Spain \\ 
and Universidad de La Laguna, Departamento de Astrof\'{i}sica, E-38206 La Laguna, Tenerife, Spain}}
\begin{document} 

\date{Accepted 2015 February 27}

\volume{449}\pagerange{2396--2404} \pubyear{2015}

\maketitle

\label{firstpage}

\begin{abstract}
To provide a quantitative cosmological context to ongoing observational work on the formation histories and location of compact massive galaxies, we locate and study a sample of exceptionally compact systems in the {\sc bolshoi} simulation, using the dark matter structural parameters from a real, compact massive galaxy (NGC 1277) as a basis for our working criteria. We find that over 80\% of objects in this nominal compact category are substructures of more massive groups or clusters, and that the probability of a given massive substructure being this compact increases significantly with the mass of the host structure; rising to $\sim30\%$ for the most massive clusters in the simulation. Tracking the main progenitors of this subsample back to $z=2$, we find them all to be distinct structures with scale radii and densities representative of the population as a whole at this epoch. What does characterise their histories, in addition to mostly becoming substructures, is that they have almost all experienced below-average mass accretion since $z=2$; a third of them barely retaining, or even losing mass during the intervening 10\,Gyr.
\end{abstract}.  
 

\section{Introduction}\label{Introduction}

Massive (M$_\star\gtsim 10^{11}\,\Msun$) compact (R$_{\rm e}\ltsim 1.5$\,kpc) galaxies are extremely rare in the present-day Universe, several groups estimating their number density to be currently $n<10^{-6}$Mpc$^{-3}$ \citep[]{Trujillo09,Taylor10}. As well as for their scarcity, these objects are of great interest because their structural properties \cite[]{Trujillo12} are comparable to a large fraction of the massive galaxy population 10 Gyr ago at a redshift of $z\sim 2$ \citep[e.g.][]{Trujillo07,Buitrago08,Carrasco10,Chevance12,Szomoru12}, when compact massive galaxies were significantly more abundant \citep[ $>10^{-4}$\,Mpc$^{-3}$;][]{Taylor10}.

As cosmic time has evolved, these galaxies have progressively disappeared \citep{Stockton10,Valentinuzzi10,Cassata11,Poggianti13,Damjanov14,Hsu14}.  The decline in the number of these galaxies has been linked with merging \cite[]{Lopez12,Newman12,Oser12,Bedorf13,Hilz13,Quilis13,Shankar13} and this scenario can explain a large number of observations \citep[e.g.][and references therein]{Trujillo11}. 

The recent discovery of a nearby compact massive `relic' galaxy, NGC 1277, with similar structural properties to typical galaxies of the same mass at z$\sim$2 \cite[]{Trujillo14}, has opened the possibility of exploring the properties of this galaxy population in unprecedented detail, including in particular some hints about the dark matter halo properties \cite[]{vandenBosch12}. Having access to these characteristics is one key to understanding this type of object in a cosmological context. In fact, we lack knowledge about the connection of these compact massive galaxies with the cosmic structures which they inhabit. In particular, which kind of dark matter halo they occupy and how these halos evolve with time.

In this paper we aim to fill this gap, and address the following questions: How do these particular dark matter structure properties compare with the rest of the population? Are they substructures of larger dark matter halos or are they isolated objects? How have the structural properties of these dark matter halos evolved with time?

We confront these questions from a cosmological perspective by taking the dark matter structure properties that would be consistent with our knowledge of this real system (\S\ref{CaseStudy}), and then looking for structures in a numerical simulation that share these properties (\S\ref{z0}). These candidate structures thus identified, we can then begin asking where they came from and what has dictated their evolution (\S\ref{z1}).

\section{Associating compact galaxies and compact structures}\label{CaseStudy}

We begin by characterising the properties of one particularly well-known compact massive galaxy, NGC 1277, described in \cite{Trujillo14} to have a stellar mass of $M_\star = 1.2 \pm 0.4 \times 10^{11} M_\odot$, half of which is contained within a circularised radius of just $r_{\rm e} = 1.2\pm 0.1$\,kpc.   

To make the association between real systems such as this, and the structures in a cold dark matter (CDM) simulation, we can draw on both empirical, semi-empirical and theoretical arguments.

Ideally, we might rely entirely on the empirical measurement of the real system's mass profile, but unfortunately the profile is only marginally constrained by direct measurement. However, we can still note the best fitting profile parameters, found from a full analysis of the stellar kinematics by \citet{vandenBosch12}. These are: $\rho_{\rm s} = 0.027\,\Msun {\rm pc}^{-3}$ and $r_{\rm s}=26$\,kpc, where these apply to the `NFW profile', the standard fitting function of \citet{Navarro95,Navarro96,Navarro97} for density as a function of radius (\ref{NFW}).

To use these figures directly to seek analogue structures in CDM-only simulations is of course to neglect the effects of the formation of the galaxy itself on the profile.  But given the already stated uncertainty in $\rho_{\rm s}$ and $r_{\rm s}$, a detailed attempt to analytically reverse-model the effects of collapse and ejection of the normal matter component would be inappropriate.  Appealing also to numerical work on this issue, a recent study by \citet[][]{Schaller14} compared the density profiles of structures in the hydrodynamical {\sc eagle} simulations \citep{Schaye15} with the equivalent structures from a `dark matter only' version. Though both $r_{\rm s}$ and $\rho_{\rm s}$ did differ between the two cases, with standard deviations at the relevant mass range of about a factor of two, the {\em expectation} change was always close to zero. There was no clear tendency for the absence, or presence, of normal matter in the numerical simulations to shift these parameters systematically one way or other.

To add to these empirical measurements of profile parameters, we can also use the stellar mass of NGC 1277 to indicate a likely structure mass range, by appealing to the abundance-matching hypothesis \citep[e.g.][]{Behroozi10,Moster10}. In particular, the expected halo mass for a given stellar mass\footnote{as opposed to the expected stellar mass for a given halo mass.}, calculated in this way by \citet{Behroozi10}, would associate our stellar mass range here, $M_\star \approx 0.8 - 1.6 \times 10^{11} M_\odot$, with halo masses of:
\begin{equation}
M_{\rm v} \approx 2\times10^{12} - 10^{13} M_\odot~, \label{mass_range} 
\end{equation}
where this approximate range incorporates both the uncertainty in $M_\star$ and the scatter in the mapping to $M_{\rm v}$.

As well as using the stellar mass as a guide, we can also draw on standard theoretical arguments which would expect the {\em radial} extent of the final galaxy to broadly reflect the quantity and distribution of angular momentum in the structure in which it formed \cite[e.g.][]{Fall80,Fall83,Mo98}. This idea has recently received a boost of empirical support from \cite{Kravtsov13}, who showed that galactic radii and host structure radii linked by the abundance-matching ansatz are directly proportional: $\langle R_{\rm e}\rangle \approx 0.015\,\langle R_{\rm 200}\rangle$ with a scatter of $\approx 0.2$ dex ($R_{\rm 200}$ being the radius that encloses 200 times the critical density).

This correlation would pair a galaxy of $R_{\rm e}\approx 1.2$\,kpc with a structure of $R_{\rm 200}\approx 80$\,kpc, which at $z=0$ corresponds to $M_{200} \approx 200 H_0^2 R_{\rm 200}^3 / 2{\rm G}\approx 6 \times10^{10}\Msun$, or a virial mass of approximately $M_{\rm v} \approx 10^{11}\Msun$. The discrepancy between this mass, and the range in the preceding paragraph deduced from the stellar mass, is a reflection of the exceptional properties of NGC 1277.  

To reconcile this, we need to go beyond the theory's application to mean or representative values. The standard, more detailed version is that the net {\em specific} angular momentum of material is conserved in the formation of the galaxy. This would mean that a more compact halo, with more of its mass at smaller radii, leads to a galaxy formed from material with less angular momentum -- and thus more compact -- than others of the same mass. 

For the purposes of our exercise here, it would therefore seem sufficient to consider a category of compact structures whose mass exceeds the lower limit of the range indicated in (\ref{mass_range}), have highly centrally concentrated mass distributions, but with limits chosen such that our real system would be comfortably included in the sample (based at least on the best fitting parameters which are available).

\begin{figure*}
\includegraphics[trim=20mm 39mm 30mm 21mm,clip,width=\textwidth]{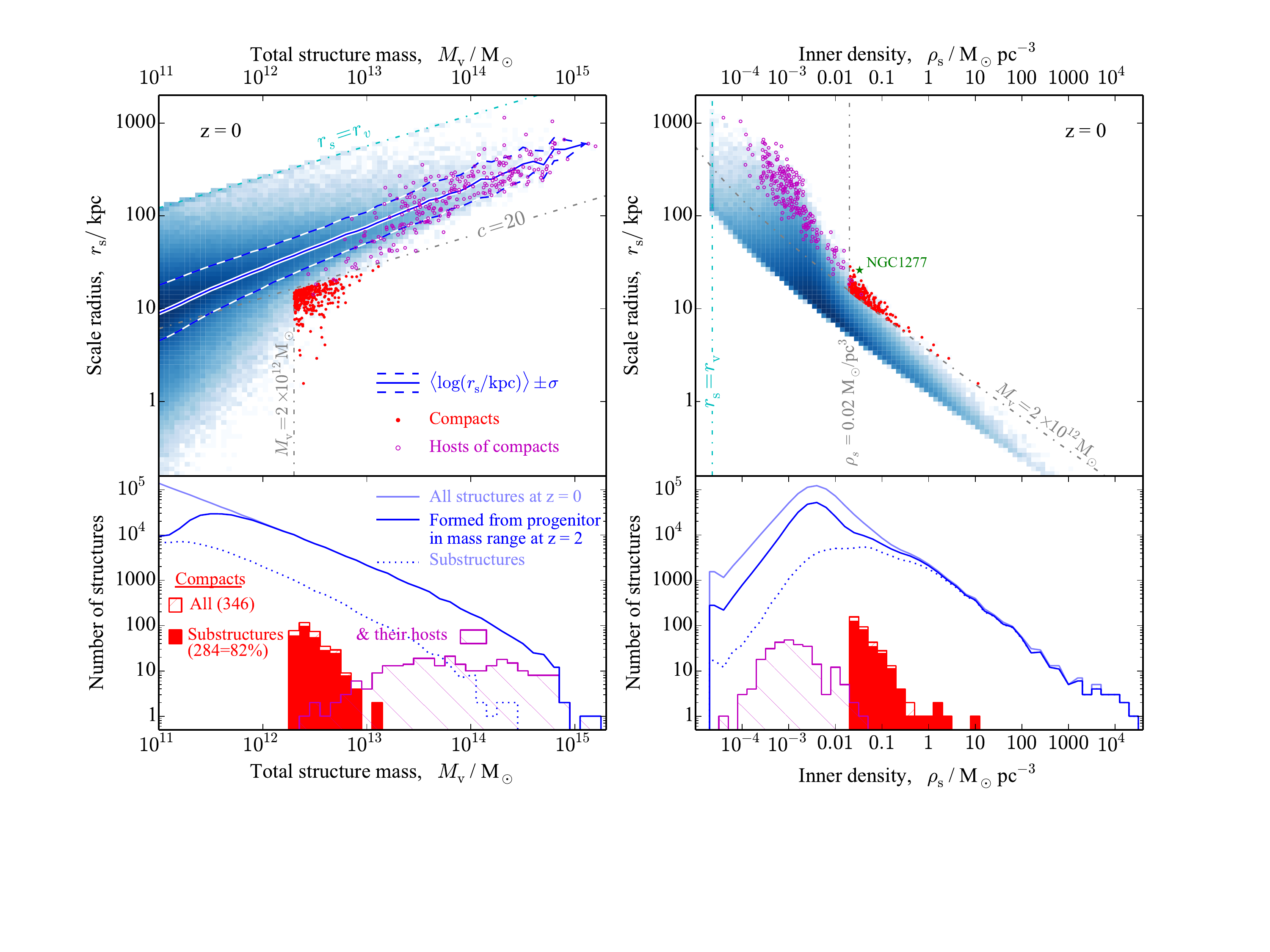}
\caption{The position of compact structures relative to the overall distributions in the {\sc bolshoi} simulation, and to the position of their host structures, where relevant. The {\bf top panels} show the the $M_{\rm v} - \rs$ plane (left) and the $\rho_{\rm s}-\rs$ plane (right) using the same key. Our category of `compacts'  are highlighted as dots, their hosts in circles, and the distribution of the population is indicated with background shading. The mass and density defining our nominal selection criteria (\ref{limits}) and other relevant loci, as labelled, are highlighted with dot-dashed lines. The star in the upper right panel indicates the best-fitting profile parameters for NGC 1277. The {\bf bottom panels} show the projected distributions on to the respective x-axes.}\label{descendants1}
\end{figure*}

We thus define our sample of interest in terms of characteristic\footnote{`Characteristic density' is the term used by \citet{Navarro97}. The mean enclosed density for their profile (\ref{NFW}) falls to this value, $\rho_{\rm s}$, at a radius $r=0.72\,r_{\rm s}$. As $r_{\rm v}\sim5-10\,r_{\rm s}$, this density $\rho_{\rm s}$ is representative of a fairly centralised region, typically just a few percent of the total volume. Hence also `inner density' \citep[][and here]{Bullock01,Wechsler02}.} or inner density, $\rho_{\rm s}$, and total virial mass, $M_{\rm v}$:
\begin{equation}
\rho_{\rm s} >0.02\,\Msun {\rm pc}^{-3} \hspace{0.8cm} {\rm and} \hspace{0.8cm} M_{\rm v} > 2\times 10^{12}\,\Msun ~,\label{limits}
\end{equation}
This total inner density value corresponds to a dark matter density of $0.02(1-\Omega_{\rm b}/\Omega_{\rm M})\approx0.016$, so the best-fitting dark matter inner density for NGC 1277 would fall within this category by about a factor of two. The sample also contains the best fitting scalelength by a similar margin, as can be seen from the upper right panel of Fig. \ref{descendants1}. In terms of concentration parameter, $c\equiv r_{\rm v}/r_{\rm s}$, the criteria in (\ref{limits}) amounts to a selection of about $c \gtsim 20$.

So, based on all these combined considerations, we have a well-motivated and interesting category of compact, massive structures. The next step is to locate such structures in a large numerical simulation and investigate their whereabouts and histories. This now follows in \S\ref{Simulation}.

\section{Analogue compact structures in simulations}\label{Simulation}

The following results and figures are taken from the publicly available catalogues of structures found within the {\sc bolshoi} simulation \citep[][and references therein]{Klypin11,Behroozi13b} using the halo finding algorithm presented in \citet{Behroozi13}. The simulation is a cubic box of 250$h^{-1}$ comoving Mpc containing $2048^3$ particles of mass just under $2\times 10^8\,\Msun$.  

Cosmological parameters were chosen to be consistent with the WMAP five year \citep{Dunkley09,Hinshaw09,Komatsu09} and seven year data \citep{Jarosik11}, notably: $h0 = 0.7,~\Omega_{\rm M}=0.27,~\Omega_{\rm b}=0.0469$, $\sigma_8=0.82$ and $n_{\rm s}=0.95$.

The virial masses, $M_{\rm v}$, and radii, $r_{\rm v}$, for structures in the simulation are defined by \citet{Klypin11} relative to the mean universal matter density at that redshift:
\begin{equation} 
{\rm G}M_{\rm v} = \nicefrac{1}{2}\Delta_{\rm v}(z)H_0^2\Omega_{\rm M}(1+z)^3r_{\rm v}^3~,\label{virial_defenition}
\end{equation}
where $\Delta_{\rm v}(z)$ is the virial overdensity. For the adopted cosmology, $\Delta_{\rm v}(0)=360$, which is equivalent to $\Omega_{\rm M}\Delta_{\rm v}(0)=97$ times the {\em critical} density.  The nominal resolution limit of $M_{\rm v}\approx 10^{10}\,\Msun$ \citep{Behroozi13}, is comfortably exceeded by all the structures that we are concerned with in this study.

\subsection{Where are they now?}\label{z0}

The final distribution of masses, radii and inner densities of structures in the simulation is shown in Fig. \ref{descendants1}, highlighting in particular the location of the 346 structures in the volume ($n\approx7\times10^{-6}{\rm Mpc}^{-3}$) that meet our nominal compact criteria defined by (\ref{limits}). This subsample constitutes the most compact 1\% of all structures in the same mass range, as can be appreciated by their location at the very bottom edge of the $M_{\rm v}-r_{\rm s}$ plane, shown in the upper left panel.

The upper right panel shows the $\rho_{\rm s}-r_{\rm s}$ plane. The best-fitting parameters for NGC 1277, indicated with a star, show that these values would place it inside our compact subsample (though the limits were motivated by a combination of factors; see \S\ref{CaseStudy}).

The lower left panel of Fig. \ref{descendants1} shows the total mass distribution of the population and the subsample, divided by distinct systems and substructures. This provides a key result that over 80\% of such compact massive systems are identified to be substructures of a larger collapsed region (as compared to 8\% of all structures in the same mass range). Furthermore, of the $\sim 20\%$ of these exceptional systems that are in fact distinct, we note that over half are found to have passed through larger structures in the past. So, in total, 94\% of the objects in this subsample are or have been influenced by a larger collapsed region.

This is in agreement with the overall results from this simulation, which showed a trend for substructures to be more compact on average than distinct systems of the same mass \citep[][eqns. 10-11]{Klypin11}, and with earlier work \citep[e.g.][]{Ghigna98,Bullock01}.  It is also in agreement with the actual location of our case study, NGC 1277, in the massive Perseus Cluster.

The hosts of these compact satellites\footnote{About 10\% of the hosts contain two of the compact systems (none contain three), so there are slightly fewer hosts than there are compact substructures.} are also highlighted in Fig. \ref{descendants1}, and are shown to have very flat distribution in mass from $10^{13}-10^{15}\,\Msun$.  So {\em given that you have a compact system} it is similarly likely to turn out to be in any sufficiently massive host. 

The clause here is emphasised because the converse statement, is {\em not} true. Taking a potential host structure at random, it is overwhelmingly {\em more likely} to find one of these compact substructures if the host is more massive. This is clear when the number of hosts-of-compacts (hatched area in Fig. \ref{descendants1}) is considered as a fraction of the number of systems in general (solid line above it), which is rapidly declining.  However this is to some extent a trivial result in the sense that there are just {\em more objects} inside the most massive clusters. Whatever it is you are looking for, it therefore seems a good bet that you are more likely to find it there, and compacts are apparently no exception.

\begin{figure}
\includegraphics[trim=1mm 2mm 12mm 79mm,clip,width=\columnwidth]{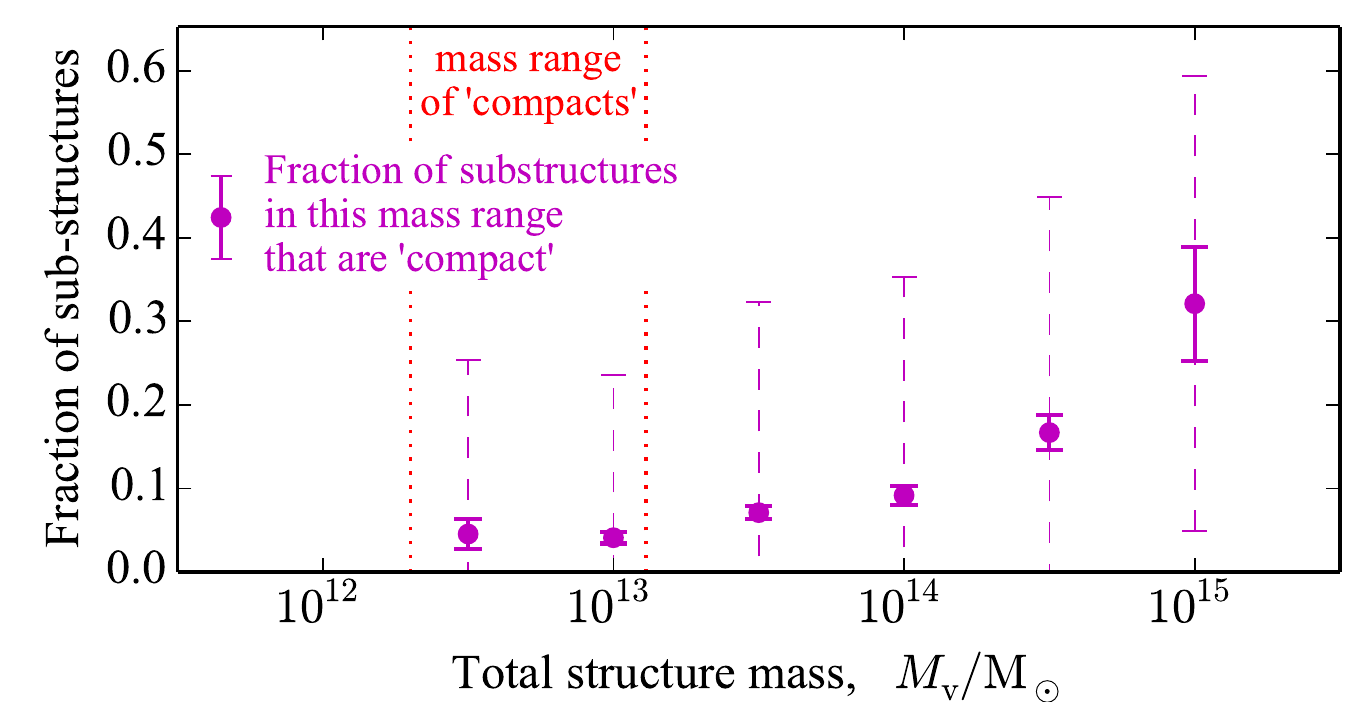}
\caption{The distribution of the compact structures in our nominal definition (highlighted in Fig. \ref{densities}) 
as a function of the mass of their host structure at $z=0$. Specifically, for each host structure we find the number of compact substructures within it and divide by the number of substructures that lie in the same mass range. Points show the mean of this fraction for all structures in each host mass bin that contain one or more substructures in the mass range. Outer, dashed error bars give the standard deviation in this fraction, and inner error bars indicate the error on the mean (i.e. $\sigma/\sqrt n$, where n is the number of host systems in the mass interval which contain substructures from the given mass range). Systems containing no substructures in the mass range are not included in the statistics.}\label{fractions}
\end{figure}

But with this large and detailed catalogue at our disposal, it should be possible to get a little beyond these two basic statistical effects and see if -- after these have been accounted for -- there remains any other evidence of a preferred environment for this category of system (i.e. one that is not also the case for systems in general). 

For those that are still substructures, we seek some normalisation of probability that gives us a fair sense of their likely whereabouts. One such normalisation is to look at their abundance as compared with other substructures that lie in the same mass range as they do. This comparison is shown in Fig. \ref{fractions}.

\begin{figure}
\includegraphics[trim=1mm 2mm 12mm 79mm,clip,width=\columnwidth]{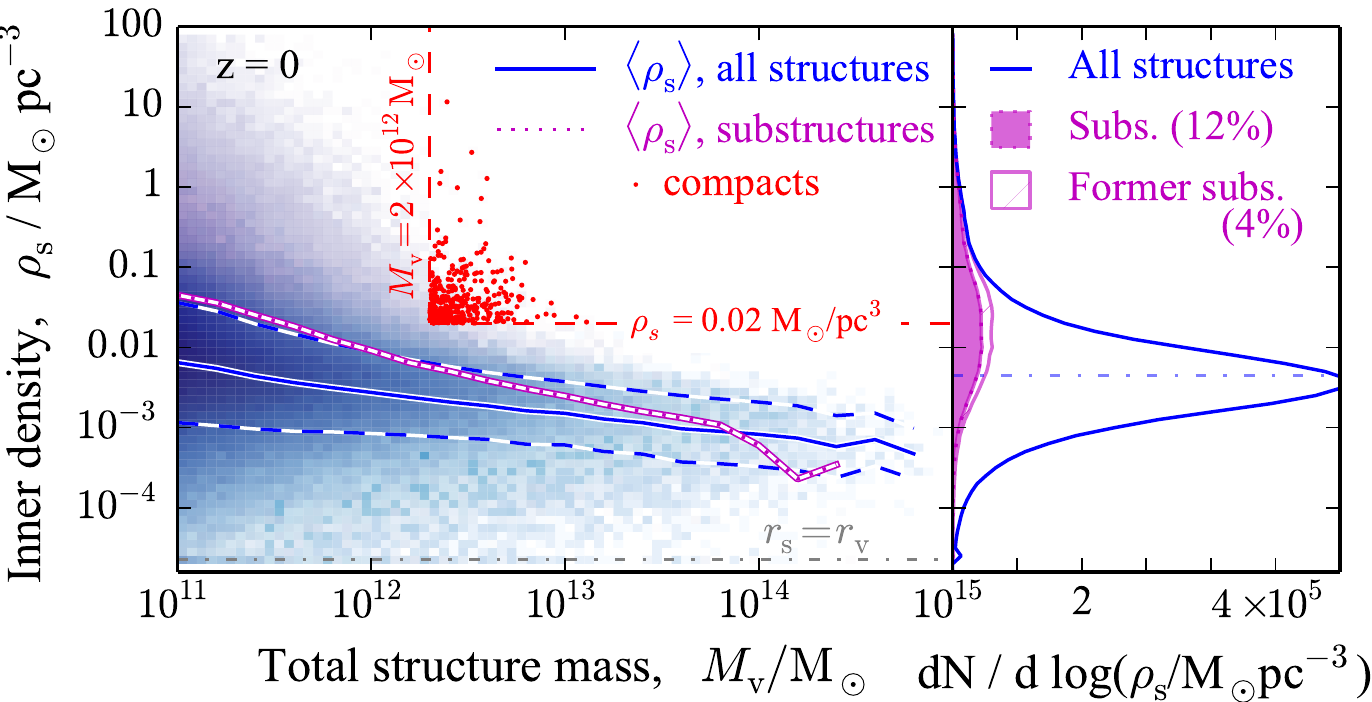}
\caption{The distribution of systems in the mass--density plane (left panel), showing also the projection on to the density axis (right panel). Solid and dashed lines show the mean and standard deviation of the inner density of all structures in each mass bin. The dotted line shows the mean for substructures only.}\label{densities}
\end{figure}

The result of this calculation is this: if one finds a substructure in the given mass range ($\sim2\times 10^{12}-10^{13}\,\Msun$), then the probability of it being from this highly compact category -- the most compact 1\% of all structures in that mass range -- increases from a few to 30\% {\em as a rising function of the mass of the host structure}. 

There is considerable scatter in the results, as indicated by the dashed error bars in Fig. \ref{fractions}. This is perhaps to be expected given the low number statistics involved in the fraction itself; the number of such massive substructures contained in any one group or cluster is at most eight. Many hosts in the sample have one or more massive substructures but no `compacts' (i.e. 0\%) and some have just one massive substructure which {\em is} compact (i.e. `100\%'). 

However, as there are over 2,600 such hosts within the simulated sample, the trend seen in Fig. \ref{fractions} 
is statistically significant. This is reinforced by the solid error bars, which show the much smaller error on the mean, $\sigma/\sqrt{n}$. Whilst a randomly selected structure from this mass range is only $\sim1\%$ likely to be so compact, a {\em substructure} from the same mass range in a $10^{14}\,\Msun$ cluster has about a $10\%$ chance, rising to $20-40\%$ in a $10^{15}\,\Msun$ cluster.

These conclusions should be followed by reiterating that these figures are specific to our choice of category, that the percentages given do not include host structures which have {\em no} massive substructures, and that 20\% of this compact category are in fact not substructures at all. But, that said, Figs. \ref{descendants1} and \ref{fractions} reveal a great deal about these systems' whereabouts in the single, $z=0$, halo catalogue. 

Before moving on to higher redshifts, there are a few small points of interest concerning features of the general distribution of structures revealed in Fig. \ref{descendants1}.  Firstly, the mean scale radius as a function of mass, in the upper left panels, is not far from tracking a constant mean density solution, but is a little steeper reflecting the fact that, at low redshifts, less massive halos tend to be more concentrated \citep[][fig. 5]{Klypin11}. A specific consequence of this shows up in the distribution of inner densities, to the right, showing that {\em all} the highest inner densities are found in small, lower mass systems. A glance at the lower right panel confirms also that all these very high inner density systems are substructures.

\begin{figure*}
\includegraphics[trim=20mm 39mm 30mm 21mm,clip,width=\textwidth]{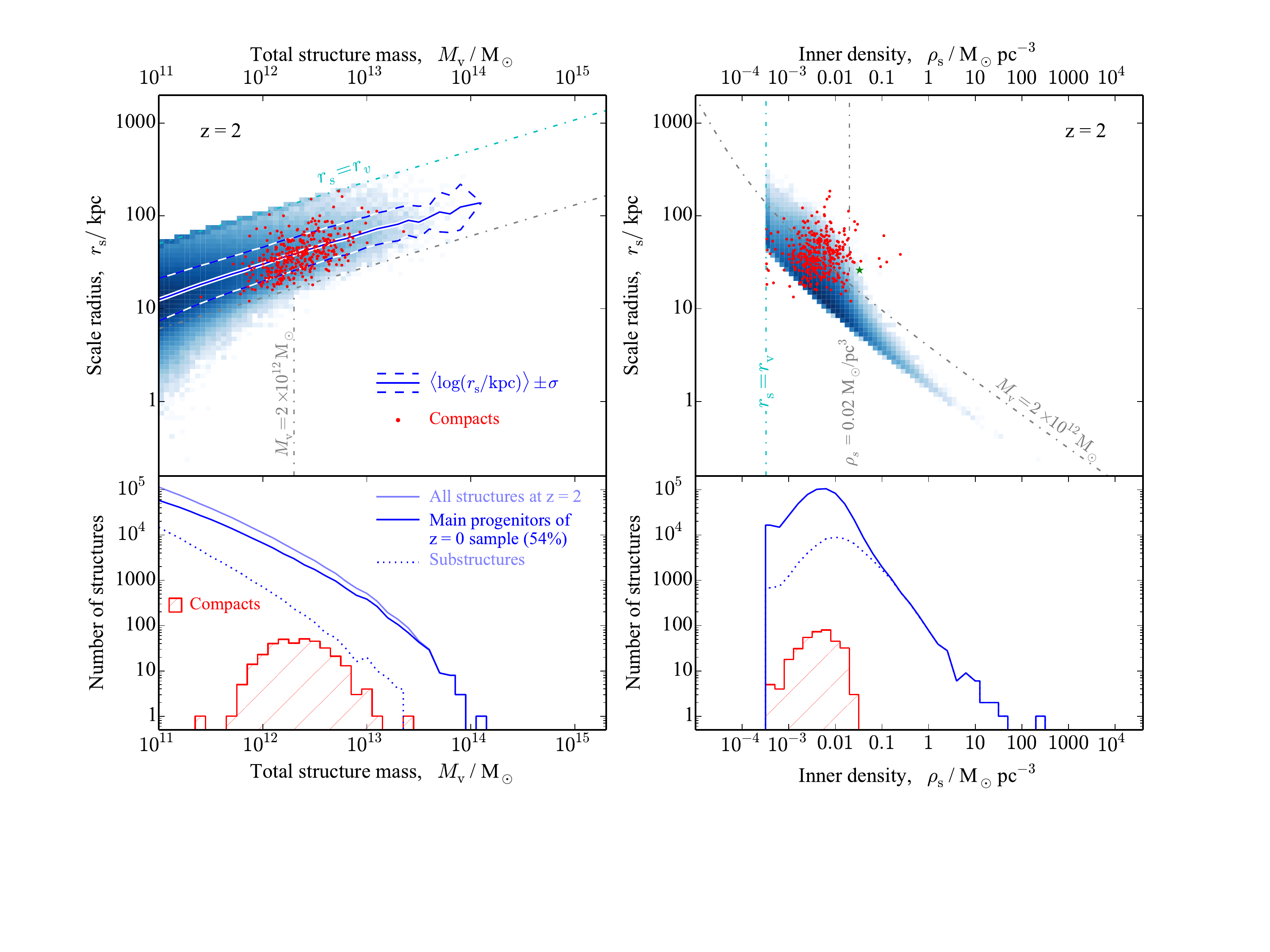}
\caption{The distributions of the structure population at $z=2$, using the same axes and key as Fig. \ref{descendants1}. The highlighted subsample here are the most massive progenitors of the systems that were highlighted at $z=0$.}\label{descendants2}
\end{figure*}

Having mentioned these features, it is worth taking a quick look at the mass--density plane itself, shown here for reference in Fig. \ref{densities}. This firstly reconfirms the simple point that all the top three orders of magnitude in inner density (e.g. $\gtsim 0.1\,\Msun/{\rm pc}^3$) occur in low-mass substructures  (i.e. $\ltsim 10^{12}\,\Msun$), or former substructures. It also shows the more general feature of decreasing inner density with total structure mass. The mean varies by a factor of ten for systems in general, with a significantly more pronounced variation of a factor of a hundred seen in substructures.

Fig. \ref{densities} also includes a projection of distribution on to the density axis, as did Fig. \ref{descendants1}, but in this case using a linear scale. This emphasises that, notwithstanding these important variations, the general distribution is quite sharply peaked and most inner densities lie within an order of magnitude of this overall mode.

\subsection{Where did they come from?}\label{z1}

Having located and characterised our subsample of compact massive structures at $z=0$, it is immediately interesting to ask what the basic features of their histories are. This should help explain what has led to them being such rare objects, or at least assess any expectations we may have had already from basic structure theory.

As a first step, we can begin by taking the same axes that were used to characterise the final properties of the structure population in Fig. \ref{descendants1}, but show the state of affairs 10 Gyr previously, at $z=2$. This is shown in Fig. \ref{descendants2}. 

Regarding first the two general distributions in these planes, these do not change a great deal from $z=2\rightarrow0$. There is, of course, continued growth of high mass structure but these new systems fall on the existing trends with scale radius and density. So the overall distributions stay the same in that sense. 

The extremes of the density distribution become increasingly populated with time. The low-density end should not be over interpreted as the NFW fits are no longer reliable as $r_{\rm s}\rightarrow r_{\rm v}$ (see appendix \ref{NFWcheck}). The high-density end, though, is interesting. That this tail extends as time goes on is perhaps not as expected, and not all these extreme objects at $z=0$ are simply older structures that have been stuck in a cluster and not evolving; many become {\em more} dense from $z=2\rightarrow0$ (see also Fig. \ref{densitychange} and Fig. \ref{densitychange_physical}).

\begin{figure}
\includegraphics[trim=1mm 2.5mm 11mm 4mm,clip,width=\columnwidth]{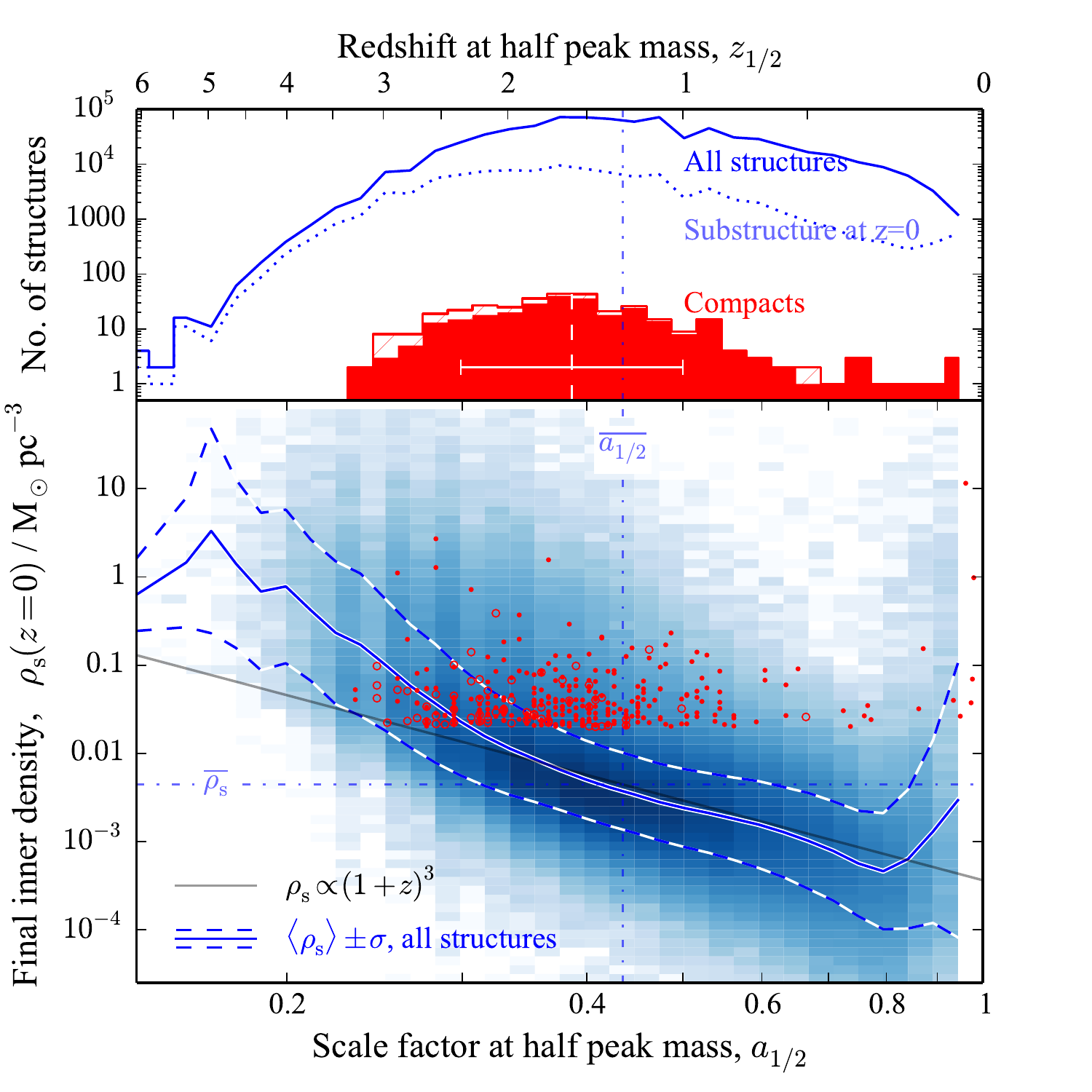}
\caption{The correlation between the final ($z=0$) inner density of structures, and the redshift or scale factor at which half their peak mass was in place. The key is as labelled, following the theme set by Figs. \ref{descendants1}-\ref{densitychange}, with the addition of a faint solid line showing the scaling relation $\rho_{\rm s}\propto(1+z)^3$, constrained to pass through the geometric means (dot-dashed lines) of both distributions.}\label{halfmass}
\end{figure}

\begin{figure}
\includegraphics[trim=1mm 2.5mm 11mm 4mm,clip,width=\columnwidth]{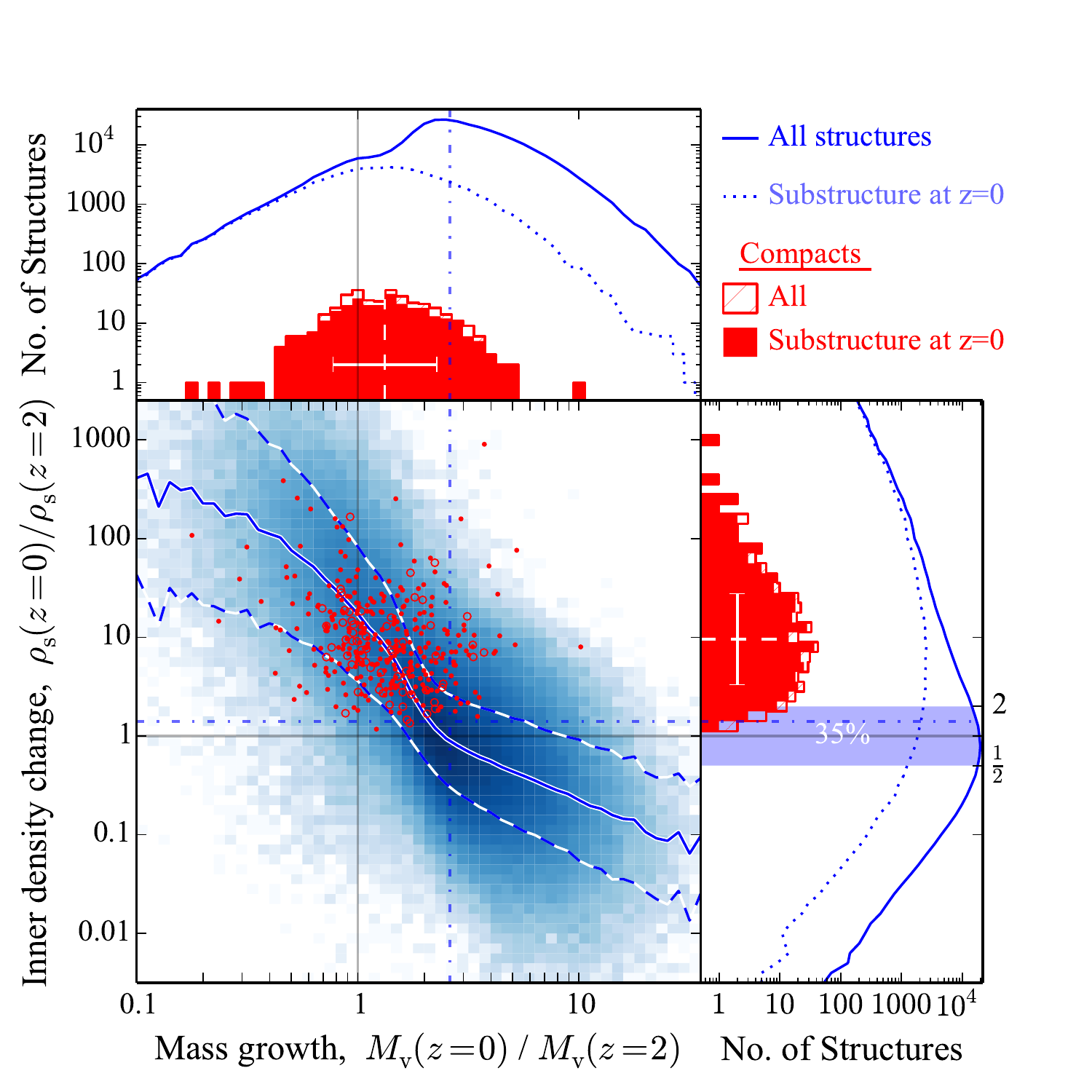}
\caption{The correlation between mass growth and inner density change, as seen in structures from the {\sc bolshoi} simulation. Background shading in the main panel shows the masses and inner densities of all substructures at $z=0$ as fractions of the same quantities for their main progenitors at $z=2$. The dot-dashed lines show the overall mean of each quantity. The solid and dashed lines show the mean and standard deviation of the density change in each mass change bin. Our category of compact structures are highlighted as circles, and as dots if they end up as substructures at $z=2$. The peripheral panels show the projection of the distributions on to the respective axes, with dotted lines showing the distribution for those which become substructures by $z=0$. The white dashed lines and error bars show the mean and standard deviation for the compact subsample. The shaded area in the right panel highlights the fact that only a minority of structures retain their $z=2$ inner density to within a factor of 2.}\label{densitychange}
\end{figure}

This is a good point to revisit our expectations concerning the relationship between the `age' of structures and their density. The familiar statement of this relationship is that structures' ``characteristic densities are just proportional to the cosmic density at the time they `formed'\,'' \citep{Navarro97}, where formation is taken to be the point at which ``half of the final mass is in collapsed progenitors more massive than 10\% of the final mass''. Here, in Fig. \ref{halfmass}, we take the opportunity to compare this statement with the much more recent simulation results we now have to hand.

Despite using the slightly later time that is recorded in the {\sc bolshoi} catalogue (the point at which half the peak mass lies in the most massive progenitor), the correlation found in Fig. \ref{halfmass} does follow this well-known rule, at least for typical formation redshifts. This can be seen by comparing the two solid lines in the bottom panel; one representing the mean half-mass redshift of structures in the simulation, and the other showing the locus $\rho_{\rm s}\propto(1+z)^3$, found by \citet{Navarro97}, normalised to go through the overall mean values here.  The two lines are overlapping from $z\approx2\rightarrow0.5$, during which period the majority ($\approx 60\%$) of all these $M_{\rm v}>10^{11}\,\Msun$ structures attain half their peak mass. 

The mean half-mass redshift for all the structures in our final $z=0$ sample turns out to be $z\approx1.3$. The mean for the compact subsample is somewhat earlier than that, but this is {\em not} necessarily what is making them exceptional. The distribution of formation times is broadly representative of the population, so we must assume that it is the later half of their evolution which makes them exceptional by $z=0$.

With this in mind we now turn back to the compact candidates' locations in Fig.  \ref{descendants2}, recalling that, at $z=0$,  these all end up in a very sparsely populated region at the edge of the distribution in Fig. \ref{descendants1}. At $z=2$, however, it seems that their most massive progenitors were evenly distributed around the mean trends in both planes. In fact, none of the progenitors fall within the region on the $M_{\rm v}-r_{\rm s}$ plane which their descendants will eventually occupy, and very few fall in this region in the $\rho_{\rm s}-r_{\rm s}$ plane.

The only thing which these main progenitors of this subsample seem to have in common at all, at this earlier time, is their mass. Though the spread is considerable (over an order of magnitude) it is not actually that much greater than the range of their descendants at $z=0$. And, crucially, the centres of both distributions are almost the same, meaning that the thing these compacts seem to have in common is {\em very little mass growth since $z=2$}.

To explore and quantify this, we can look specifically at the distribution of structures in terms of this property -- mass growth -- and see how this correlates with changes in inner density. This is shown in Fig. \ref{densitychange}. 

Concerning first the properties of the population in general, this figure shows a clear anti-correlation between mass change and inner density change.  As inner density for a given mass increases monotonically with concentration (see Appendix \ref{PhysicalDensity}), Fig. \ref{densitychange} can be interpreted also in this sense; lesser mass growth tending to lead to a more concentrated final structure, and mass {\em loss} even more so. 

Fig \ref{densitychange} also shows that the modal\footnote{The {\em mean} density change is a little above zero, due to the skew in the distribution from systems becoming substructures, but the mode is still close to zero.} density change is close to zero. This runs alongside the prominent idea of inside-out structure growth \citep{Diemand07,Cuesta08}, which can be further corroborated by looking also at the changing density inside a fixed physical radius, shown for reference in Appendix \ref{PhysicalDensity}. 

But, in particular when mentioning that notion, the scatter in all directions in this figure cannot be emphasised enough. All four quadrants of the main panel are occupied. It is fair to say that the true fraction of the population lying away from the origin is exaggerated somewhat by the logarithmic scales, but we set this visualisation issue aside by including the simple statistic in the right panel that {\em a clear majority of structures experience an inner density change of more than a factor of 2 from $z=2\rightarrow0$}.

So in this simulated universe, at least, most outcomes are possible. Some consistency with our general rules for structure formation can be seen, but they are indeed {\em general}; we should not be surprised when any one particular bit of universe -- real or simulated -- chooses not play by them.

This brings us back to the particular bits of the simulation that constitute our subsample of compacts. As might be expected from the original selection criteria, they are all in the upper half of the main panel, all experiencing an increase in inner density between the two epochs. This increase is in most cases very significant, averaging over a factor of 10 for the subsample as a whole, and in several cases exceeding a factor of 100.

Fig. \ref{densitychange} also confirms the feature identified qualitatively in Fig. \ref{descendants2}, that most of the subsample change relatively little in mass ($\langle\log[M_{\rm v}(z=0)/M_{\rm v}(z=2)]\rangle=0.12$) and some 30\% actually decrease. At the same time, about 10\% of the subsample exceed the average mass increase for the population, thus reinforcing again the point that we should expect exceptions.

\vspace{3cm}

\section{Summary}

This study begins by taking the parameters of a known massive, compact galaxy and looking for simulated structures in the {\sc bolshoi} CDM cosmological simulation that would broadly correspond with these properties. We find that the best-fitting inner density and NFW scale radius of the real system would indeed put it amongst the most compact 1\% of structures in the simulated population from the relevant mass-range (Fig. \ref{descendants1}).

Of these analogue compact structures in the simulation, the great majority ($\gtsim 80\%$) are substructures within larger collapsed regions at $z=0$, and the majority of the rest were substructures in the past. The hosts at $z=0$ are evenly distributed in mass, meaning -- due to decreasing numbers overall -- that the probability of finding such a compact structure in a given host increases with the host's mass (Fig. \ref{fractions})

In terms of the {\em probability of a substructure of this mass being compact} -- which is perhaps a more relevant statistic -- we find that this also increases with host mass, the mean rising from $5\rightarrow30\%$ as $M_{\rm host}$ increases from $10^{13}\rightarrow10^{15}\,\Msun$.

We then trace the main progenitors of these compact analogues back to $z=2$, where they are all found to be distinct systems. At this earlier time, they also had unremarkable profile parameters (Fig. \ref{descendants2}), none of which would fall into the extreme category that would later define them at $z=0$. What the subsample did appear to have in common was that they are mostly:
\begin{itemize}
\item{distinct systems that become substructures,}
\item{increasing relatively little, or decreasing, in total mass}.
\end{itemize}

The latter conjecture was quantified, finding that the average increase in total virial mass to be just half the average growth for the population (Fig. \ref{densitychange}). Their inner densities, meanwhile, had all increased; on average by a factor of 10, and many by over 100.

Concerning the population of simulated structures as a whole, we find a clear anti-correlation in the simulation between $\Delta M_{\rm v}$ and $\Delta\rho_{\rm s}$; mass {\em loss} usually leading to higher inner densities (more centrally concentrated mass), and late mass accretion leading on average to lower eventual inner densities (less centrally concentrated).

In conclusion, the evolutionary paths of simulated structures revealed even just by this simple analysis, are much more rich and varied than our stock phrases on structure formation would encapsulate. Even the relatively small volumes of our simulated universes are now big enough that improbable things exist, and in great numbers. So we should not be afraid to seek them in the real one.

\section*{Acknowledgements}
All the authors acknowledge financial support from The Spanish Ministry of Economy and Competitiveness (MINECO) under the Severo Ochoa Program: Martin Stringer from grant MINECO SEV-2011-0187, Ignacio Trujillo from grant AYA2013-48226-C3-1-P and Claudio Dalla Vecchia and Inma Martinez-Valpuesta from grant AYA2013-46886-P. This research was also supported in part by the National Science Foundation under grant nos. PHYS-1066293 and PHY11-25915, a grant from the Simons Foundation, and the hospitality of the Aspen Center for Physics. Thanks go to Joel Primack and Peter Behroozi for their encouragement and assistance with the {\sc bolshoi} and {\sc rockstar} results, to Matthieu Schaller for his help interpreting the results from {\sc eagle}, and to Remco Van den Bosch for the observational data on NGC 1277. Finally, we would like to thank the referee for their insightful and constructive review.

\bibliographystyle{mn2e_Daly}
\bibliography{compacts_references}

\appendix
\section{Checking the density profiles}\label{NFWcheck}

The inner densities referred to in all the figures and analysis in \S\S\ref{z0}-\ref{z1} are taken from the best-fitting NFW profile to each structure:
\begin{equation}
\rho(r) = \frac{\rho_{\rm s}}{\frac{r}{r_{\rm s}}\left(1+\frac{r}{r_{\rm s}}\right)^2}\label{NFW}
\end{equation}
The profiles are constrained to satisfy $M_{\rm NFW}(r_{\rm v}) = M_{\rm v}$ and then the best-fitting value of $r_{\rm s}$ (given this constraint) is obtained. The inner density is therefore found from:
\begin{equation}
\rho_{\rm s} = \frac{M_{\rm v}}{4\pi r_{\rm s}^3}\cdot \frac{1}{\ln\left(1+c\right) - c/\left(1+c\right)} \hspace{1cm} \left[c\equiv\frac{r_{\rm v}}{r_{\rm s}}\right] \label{rho_s}
\end{equation}
For the analysis to carry weight, it is therefore important to check whether or not these fits were successful.

One way of assessing this is to find the radius which encloses 2500 times the critical density, $r_{2500}$, and compute the mass predicted to lie within this radius according to the best fitting NFW profile. Dividing this expectation by the {\em actual} mass enclosed, $M_{2500}$, then gives a useful measure of the fit accuracy (i.e. a value of one suggests a good fit).

\begin{figure}
\includegraphics[trim=15mm 157mm 80mm 25mm,clip,width=\columnwidth]{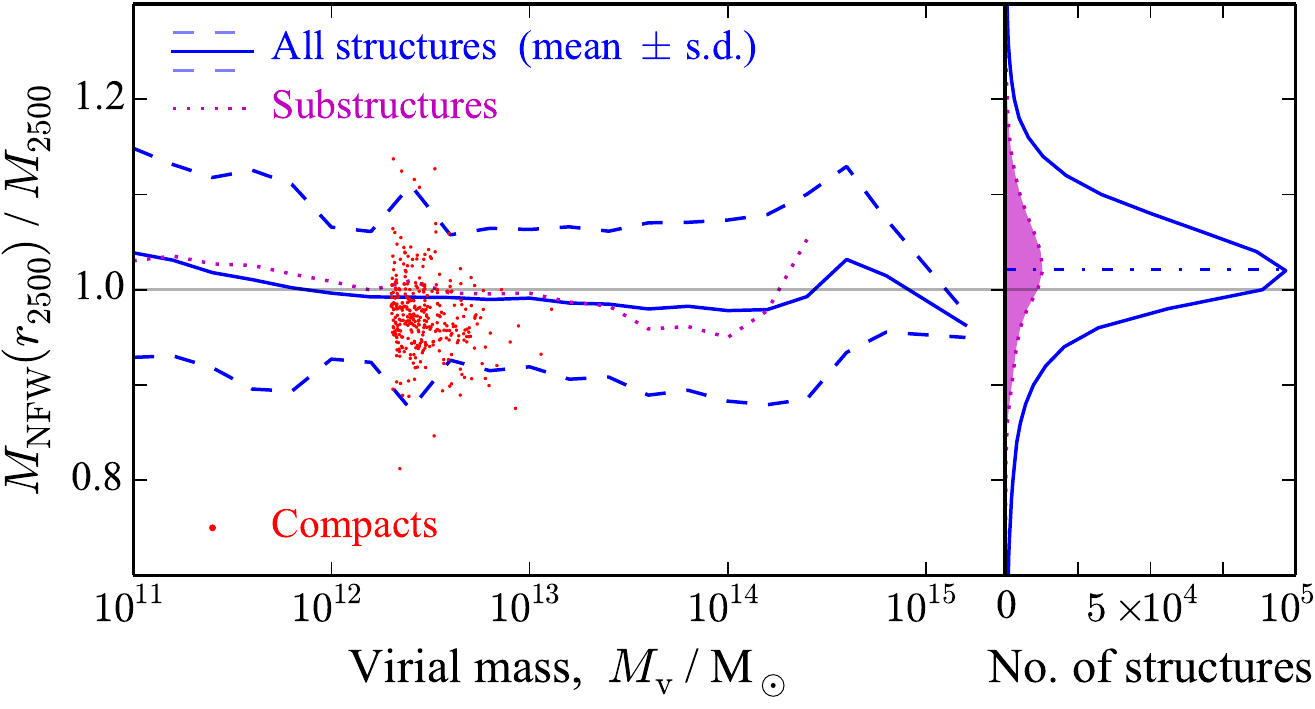}
\caption{An indication of the accuracy of the NFW density profile fit to the structures in the {\sc bolshoi} simulation, using the ratio of mass enclosed within  $r_{2500}$ according to the fit and the actual mass enclosed. The main panel shows the variation in the mean and standard deviation of this ratio as a function of structure mass, and also the mean value for substructures. The compact subsample are highlighted as dots. The right panel shows the projected distribution in the ratio for the entire mass range.}\label{NFWtest}
\end{figure}

The result of this simple assessment for the population of structures at $z=0$ is presented in Fig. \ref{NFWtest}, showing that the enclosed mass predicted by the NFW fit is usually within 10\% of the true value, and almost always within $\sim 20\%$. More quantitatively, 98.6\% of the structures have a sufficiently good fit to lie within the limits of the $y$-axis in Fig. \ref{NFWtest}; as can be appreciated by looking at the projected distribution in the right panel.

The compact structures we are particularly interested in are also representative of this general distribution. The majority of fits to these profiles lead to very nearly the correct value for  $M_{2500}$, and even the few poorer fits are still within 20\% of the mark.

Regarding the one or two in every hundred structures that have significantly different profiles (i.e. lie outside the plot range), these are evenly distributed with mass and there is no tendency for being either distinct or substructures. What these `non-NFW' systems do have in common is that they nearly all produce fits with larger scalelengths, $r_{\rm s}\rightarrow r_{\rm v}$. 

\begin{figure}
\includegraphics[trim=16mm 13mm 80mm 168mm,clip,width=\columnwidth]{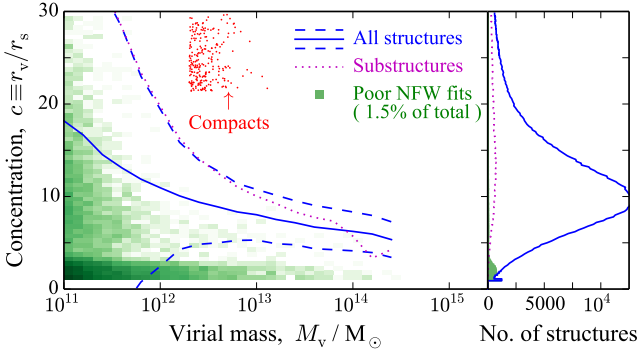}
\caption{The concentration parameters of simulated structures as a function of mass, highlighting in particular (shading) those whose profiles are poorly matched by an NFW profile, i.e. $\left| M_{\rm NFW}(r_{2500})/M_{2500} - 1 \right|>0.3$. As in the other figures, solid and dashed lines show the mean and standard deviation of the entire sample, and the overall distributions are projected on to the right panel. Our subsample of compact systems are also highlighted as dots, showing that they are safely at the opposite extreme of the distribution from the poor profile fits.}\label{concentrations}
\end{figure}

This is shown in Fig. \ref{concentrations}, in which it is also interesting to see the variation in concentration for the entire sample. The projection in the right panel confirms that almost all the poorly fitted profiles are low-concentration systems, approaching or actually hitting\footnote{About 1000 structures in this $M_{\rm v}>10^{11}\,\Msun$ sample hit this limit (i.e. $c<1.1$). Of the whole sample, this is barely $\nicefrac{1}{1000}$th, so no great concern. But of the poor profile fits in general, it constitutes about $\nicefrac{1}{10}$th.} the limit $r_{\rm s} \leq r_{\rm v}$ imposed by the fitting function.

The existence of this limit also means that the inner densities estimated using these profiles will have a forced minimum, slightly greater than the overall virial overdensity:
\begin{equation}
\rho_{\rm 0, min} = \frac{1}{\ln 2 - \nicefrac{1}{2}}\cdot\frac{M_{\rm v}}{4\pi r_{\rm v}^3}  = 1.73\,\bar{\rho}_{\rm v}(z)~,
\end{equation}
This can be seen as the left-hand cutoff in the density distributions of Figs. \ref{descendants1} and \ref{descendants2} (right-hand panels). The cutoff noticeably occurs at lower density at the lower redshift, due to evolving $\bar{\rho}_{\rm v}(z)$.

\section{Scale density and enclosed density}\label{PhysicalDensity}

Whether any given aspect of structure growth is best understood in terms of comoving distances, fixed physical distances, or some other distance, such as $r_{\rm s}$, that reflects the self-similarity of the structures, is not always obvious. The issue can be introduced by rewriting the expression (\ref{rho_s}) for the inner density of the NFW profile, in terms of $\rho_{\rm v}$, the mean density enclosed within $r_{\rm v}$:
\begin{equation}
\rho_{\rm s} = \rho_{\rm v}\cdot\frac{3c^3}{\ln\left(1+c\right) - c/\left(1+c\right)}\hspace{1cm}= \rho_{\rm v}f(c)\label{fc}
\end{equation}
i.e. the inner density is a function only of the cosmology and of $c$. So Fig. \ref{densitychange} can also be understood as showing {\em concentration} change as a function of mass growth.

\begin{figure}
\includegraphics[trim=1mm 2.5mm 11mm 14mm,clip,width=\columnwidth]{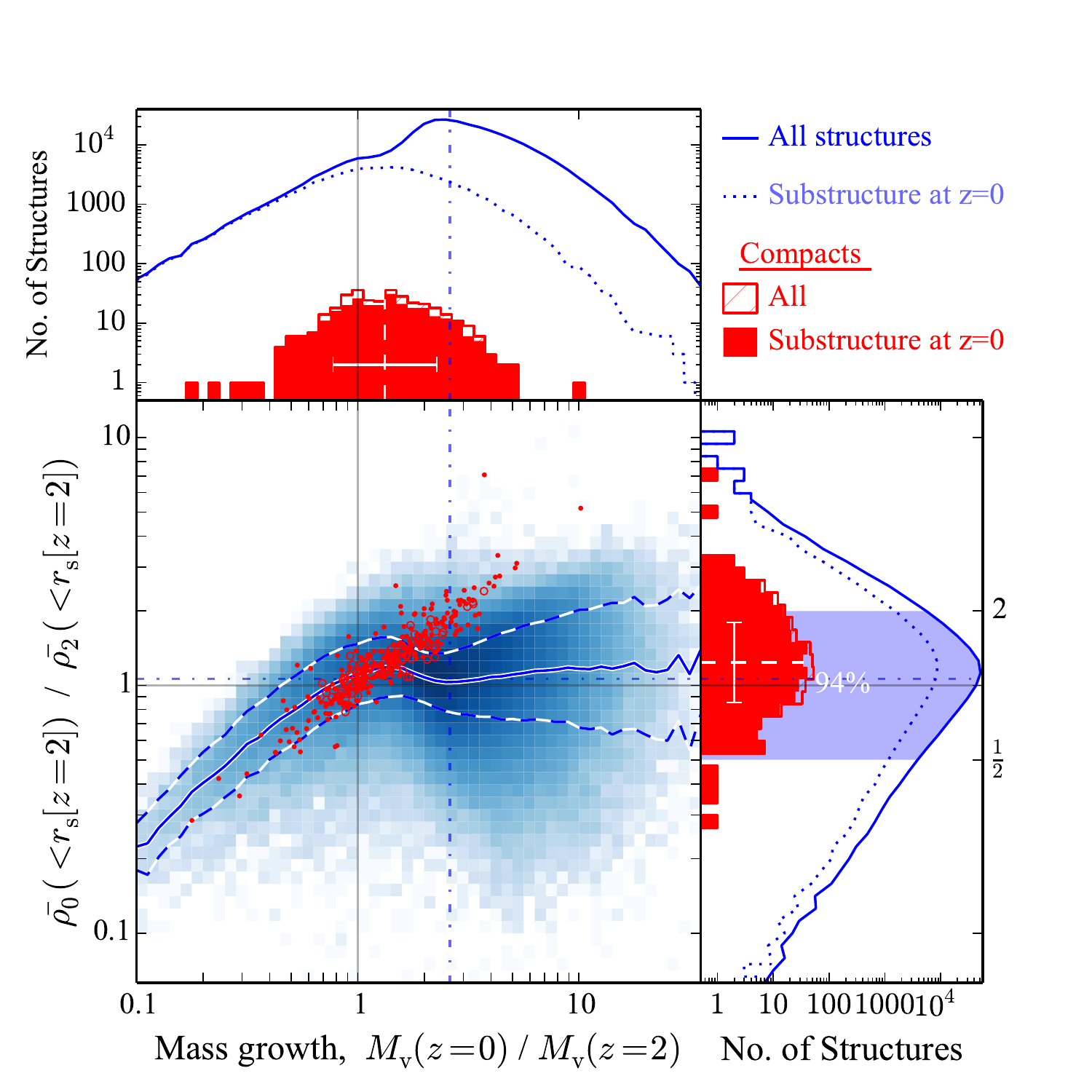}
\caption{An analogous figure to Fig. \ref{densitychange}, using an identical key but showing the change within a fixed physical radius for each structure, $r_{\rm s}[z=2]$ (rather than the change in scale density, as before).}\label{densitychange_physical}
\end{figure}

To help further interpret Fig. 6, we can investigate what is happening within a fixed physical radius in each structure as it forms, the density enclosed by fixed radius, $r$, being given by:
\begin{equation}
\bar{\rho}(<{\rm r}) = \frac{3M_{\rm v}}{4\pi r^3}\frac{f_{\rm NFW}\left(r/r_{\rm s}\right)}{f_{\rm NFW}(c)}~,
\end{equation}
where $f_{\rm NFW}(x)$ is the function: $\ln(1+x)-1/(1+x)$. The change in density enclosed by the {\em original} scale length of the halo, $r_{\rm s}[z]$, is therefore:
\begin{equation}
\frac{\overline{\rho_0}(<r_{\rm s}[0])}{\overline{\rho_z}(<r_{\rm s}[z])} = \frac{M_{\rm v}(0)}{M_{\rm v}(z)}\frac{f_{\rm NFW}(c[z])}{f_{\rm NFW}(c[0])}\frac{f_{\rm NFW}(r_{\rm s}[z]/r_{\rm s}[0])}{f_{\rm NFW}(1)}
\end{equation}
The distribution of this ratio is shown in Fig. \ref{densitychange_physical}. The key result is that mass growth does -- on average -- not affect the inner profiles of the population, over 90\% of structures preserving the enclosed density within their original scale radius to within a factor of 2. 

Our selection criteria, picking out extremely concentrated objects at $z=0$, corresponds here to a subsample which {\em do} have a slight positive mean fixed density increase. But the relative lack of growth in mass -- as discussed in \S\ref{Simulation} -- sets them apart more significantly from the population as a whole.

\label{lastpage}

\end{document}